# Computationally Supported Neutron Scattering Study of Natural Shungite, Anthraxolite, and Synthetic Carbon Black

K. Hołderna-Natkaniec[1], E.F. Sheka*[2], I. Natkaniec[1], Ye.A. Golubev[3], N.N. Rozhkova[4], V.V. Kim[5], N.A. Popova[2], V.A. Popova[2]

[1] *Faculty of Physics, Adam Mickiewicz University, ul. Umultowska, 85, 61-614 Poznań, Poland*
[2] *Peoples' Friendship University (RUDN University) of Russia, ul. Miklukho-Maklaya, 6, 117198 Moscow, Russia*
[3] *Institute of Geology, Komi Science Center, Ural Branch of RAS, ul. Pervomayskaya, 54, 167982, Syktyvkar, Russia*
[4] *Institute of Geology, Karelian Research Centre RAS, ul. Pushkinskaya, 11, 185910 Petrozavodsk, Russia*
[5] *Institute of Problems of Chemical Physics RAS, pr. Academician Semenov, 1, 142432 Chernogolovka, Russia*

*E-mail: sheka@icp.ac.ru

**ABSTRACT.** A set of $sp^2$ amorphous carbons involving natural mineral shungite carbon and antraxolite as well as two synthetic carbon blacks were investigated by using neutron powder diffraction and inelastic neutron scattering at low temperature. NDP revealed nanographite-like structure of all the samples, stacks of which are formed by basic structure units representing framed graphene molecules of ~2.5 nm in lateral dimension. INS study showed the presence of hydrogen atoms in the BSU framing area as well as of adsorbed water in the samples pores. Simulated INS spectra of adsorbed water showed its mono-layer disposition within the pores of the studied amorphics. Due to BSUs radical character, their INS spectra were simulated in the framework of both spin-nondependent (DFT) and spin-dependent (UHF) molecular dynamics. The obtained results allowed suggesting a specific INS classification of $sp^2$ amorphous carbons with respect to their hydrogeneousness based on "H-standard" INS spectra, on the one side, and riding-stimulated spectrum of heavy atoms, on the other.

# I. INTRODUCTION

Exceptionally rich allotropy is one of the outstanding characteristics of carbon, one niche of which is occupied by amorphous carbons (ACs), or carbon amorphics. Seemingly simple at first glance, the niche covers an extended set of species, both natural and synthetic, among which the former part is rather scarce and is usually presented by soots and coals, including charcoal, while the latter is quite rich including black, glassy, and activated carbons, carbon nanofoams, carbid-derived carbon, carbon fibers, diamond-like amorphous carbon ($\alpha$-C:H and (ta)-C carbons), and others (see reviews [1-9] and references therein). The list of synthetic amorphous carbons is constantly increased since more and more techniques appear to produce this highly requested material. Evidently, it should be complemented with 'technical graphenes' that cover a large class of reduced graphene oxides [10]. As for the list of natural amorphics, in view of physicists and chemists, it has been unchanged for a long time until scrupulous studies of shungite carbon [11, 12] and anthraxolite [13] have convincingly pointed that these two mineral products should be added to the list. Additionally, diamond-like natural ACs have become top issues of the carbon mineralogy of the last two decades [14, 15] as well.

The natural amorphics are products of the activity of natural laboratory during geological billion (shungite) - million (anthraxolite) years of life. Once varied as any other geological minerals, the species are nevertheless well determined and posses a number of rather standard physical and chemical properties. Synthetic amorphics are industrially manufactured in the form of hundreds of defined commercial grades that vary in their primary particle size, aggregate size, shape, porosity, surface area, and so on, just providing a very large scale of their physical and chemical properties. On this background, natural amorphics look like well characterized references, a comparison with which may threw light to the microscopic features of the synthetic ones. The current study is aimed at showing that such a comparison may be effective and fruitful. Three types of carbon amorphics related to $sp^2$-carbons of the highest carbonization have been chosen to achieve the goal: two natural ones, namely, shungite carbon (ShC) and anthroxalite (AntX), and carbon black (CB) as one of the most widely used synthetic AC products.

Thermal neutron scattering, once not quite appropriate at first glance due to the fact that carbon atoms are poor scatterers, is suitable since the selected ACs are not really 'true carbons', in contrast to, say, diamond or graphite, since their carbon content is not 100% and varies for the most carbonized species in the range of 82-95 wt % [16]. The remaining contributions are mainly provided by such light elements as hydrogen, oxygen, nitrogen, and sulfur (see last reviews [8, 16] and references therein), among which neutron scattering is highly sensitive for hydrogen-protium content. Concerning the current study, a comparative analysis of the neutron diffraction and inelastic neutron scattering spectra, supplemented by calculations, has allowed exhibiting a similar graphitic structure of the selected species and determining details of its basic structural units (BSUs).

The paper is organized as following. Section II presents methods in use exhibiting studied samples and the used facilities at IBR 2 reactor of the Joint Institute for Nuclear Research at Dubna as well as computational treatment of INS spectra. Sections III is devoted to neutron diffraction and inelastic neutron scattering experiments. The obtained data are discussed in Section IV along with the results of computational experiment. Conclusion presents the main essentials resulted from the study performed.

# II. METHODS IN USE

**Sample Preparation**

$sp^2$ ACs are usually presented as graphitic carbons with a characteristic graphitic structural pattern [17]. Such a structure of shungite carbon (ShC) has been reliably confirmed by X Ray, neutron scattering, and HRTEM studies [10, 18-22]. Similarly, HRTEM and Raman scattering have proved graphitic structure of anthraxolites (AntXs) [22]. Among synthetic amorphics, the graphitic pattern of the structure is characteristic to carbon blacks (CB) [1, 3, 6-8]. Taking this issues as leitmotiv for the sample selection, four samples were chosen, all the four belonging to the highest-carbon-content species: ShC from Shunga deposit (Karelia, Russia) [12]; AntX from Pavlov deposit (Novaya Zemlya, Russia) [22], and two carbon blacks presented by synthetic adsorbents 999632 and 999624 of Merck-Sigma-Aldrich Co [23], marked as CB632 and CB624 below. ShC and AntX were produced from the hand-picked samples from the vein of open deposit by grinding to powder with particle size less than 40 µm. ShC was additionally cleaned by repeated several successive treatments in water with stirring and followed by filtration at room temperature. After drying and rubbing homogeneous macroscopic powder with particles less than 5 µm was obtained. CB632 and CB624 were provided by producers with the indication of carbon content not less than 99.95 wt% and mean pore diameter of 6.4 nm and 13.7 nm, respectively. The samples are small-grained, consisting of 10 nm – 10 µ conglomerates. Since ShC and AntX are porous as well [11-13], all the samples may adsorb water due to which four *as prepared* samples (conditionally, *wet* samples) discussed above were complemented by four others subjected to prolong heating (drying) at moderate temperature 110-150$^0$C providing a set of *dry* samples.

**Neutron Scattering Experiments**

Neutron scattering study was performed at the high flux pulsed IBR-2 reactor of the Frank Laboratory of Neutron Physics of JINR by using NERA spectrometer [24]. The investigated samples are illuminated by white neutron beam and analyzed by time-of-flight (TOF) method on the 110 m flight path from the IBR-2 moderator. The inverted-geometry spectrometer NERA allows simultaneous recording of both neutron powder diffraction (NPD) and inelastic neutron scattering (INS) spectra. The NPD spectra were obtained for the lowest scattering angle of diffraction detectors of the NERA spectrometer, which allow to measure interplanar distances in the range of 1- 9 Å. The INS spectra were registered at final energy of scattered neutrons fixed by beryllium filters and crystal analyzers at $E_f$ = 4.65 meV at fifteen scattering angle from $20^o$ to $160^o$ by step of $10^o$. Experiments were performed at 6K and 20K.

**Computational Details**

Inelastic neutron scattering from amorphous solids, consisted of both coherently and incoherently scattering nuclea, is usually considered in the one-phonon incoherent approximation

$$\sigma_1^{inc}(E_i, E_f, \varphi, T) = O_1^{inc}(E_i, E_f, \varphi, T) \frac{\sum_k F_k(\omega)}{1 - exp\left(-\frac{\hbar\omega}{k_B T}\right)} \quad (1)$$

where
$$F_k(\omega) = \sum_{n_k} \frac{\left(b_{n_k}^{c,ic}\right)^2}{M_{n_k}} exp\left(-2W_{n_k}\right) G_{n_k}(\omega) \quad (2)$$

and
$$G_{n_k}(\omega) = \sum_j \left[A_j^{n_k}(\omega)\right]^2 \delta(\omega - \omega_j). \quad (3)$$

Here $F_k(\omega)$ is generalized vibration density of states (GVDOS) weighted with squared lengths of neutron scattering $b_{n_k}^{c,ic}$ (either coherent or incoherent) of the *k-th* nucleus as well as with amplitude of this nucleus (atom) displacement $A_j^{n_k}(\omega)$ in the *j-th* phonon mode. Other details are given in [25]. Experimental GVDOS are evaluated by attributing Ex.(1) to experimental INS spectra. Virtual GVDOS is obtained by solving one-phonon harmonic dynamic problem for modeled species. The two values comparison allows exhibiting a clear role of the *k-th* group nuclea in the studied dynamic problem.

In the current study calculations of the virtual GVDOS-$F(\omega)$ spectra were performed in the frame of both spin-nondependent (molecular density functional theory) and spin-dependent (unrestricted Hartree-Fock approximation) molecular dnamics. In the former case, the DFT generalized gradient approximation (GGA) of the exchange-correlation functional was applied, employing the Perdew-Burke-Ernzerhof (PBE) functional [26, 27] revised by Hammer (rPBE) [28]. The rPBE functional, also known as 'hard' GGA, was originally introduced to improve the adsorption energetics, being out of interest of the vibrational spectroscopy community. We have recently shown its credibility for the INS vibrational spectra calculations [25]. Molecular modeling using DMOL3 program [29, 30] and DND numerical basis set, generally comparable with the standard 6-31G(d) level. The UHF calculations were carried out by using CLUSTER-Z1 program [31], implementing semiempirical AM1 version of the tool. Both sets of computational tasks were performed at 'fine' electronic and convergence criteria as defined in Materials Studio [32]. The precisely optimized equilibrium geometries were followed with the vibrational analysis performed using the finite displacement method and completed by the GVDOS calculation using a-CLIMAX program [33]. The obtained set of δ-functions related to 3N-6 frequency modes were then Gauss-convoluted when taking into account the resolution function of the NERA spectrometer by using program NERA [34]. Thus introduced broadening takes into account multiple different internal and external factors, such as finite lifetime and anharmonicity of vibrations as well as various structural and dynamic inhomogeneities. As shown previously [25], the half-width of 80 cm$^{-1}$ is quite enough to cover these effects.

## III. EXPERIMENTAL RESULTS

**Neutron Diffraction**

$sp^2$Amorphous carbons belong to a specific class of amorphous solids made of platelets. The platelets, or basic structure units, represent graphene molecules, stacks of which exhibit short-range order followed with irregular distribution of the stacks. The ACs stack structure evidently predetermines a graphite-like image of their diffraction. Actually, Fig.1a presents a panoramic view of a set of NPD plottings for the studied *as prepared* four samples alongside with that related to spectral graphite at scattering angle $2\Theta=117.4^0$. As seen in the figure, a general NPD pattern, concerning (hkl) peak structure, of all the samples is similar to that of spectral graphite, while drastically differs from the latter by shape. As typical to graphite-like diffraction, the main features of all the amorphics NDP plottings are related to the Gr(002) reflexes located at the region of 3.3 – 3.5 Å thus pointing to undoubted graphite-like stacking of the relevant BSUs. The Gr(002) peaks are upshifted and considerably broadened pointing convincingly to limiting the stacks in size. This shifting and broadening are well pronounced and similar to ShC, AntX and CB632, while are much less for CB624 just exhibiting increasing the size of the short-range ordering of its structure. Due to high porosity, all the samples expectedly contain adsorbed water retained in the pores. The latter usually is almost completely removed by continuous heating at $110^0$ C during several hours. Dried (*dry*) samples discussed through over the paper below were subjected to such thermal treating during three days. As occurred, the water does not influence markedly NDP plottings that are identical for both as prepared (*wet*) and *dry* samples.

As seen in Fig. 1b, the narrow peak Gr(002) of graphite, the shape and width of which correspond to the resolution function of spectrometer and whose position determines $d_{002}$ interfacial distance between the neighboring graphene layers of 3.35 Å is substituted with broad peaks for the ACs. The broadening is mainly attributed to a narrowing of the coherent scattering regions (CSR) of a scatterer. According to widely used Scherrer's equation, the FWHM of a diffraction peak *B* and the

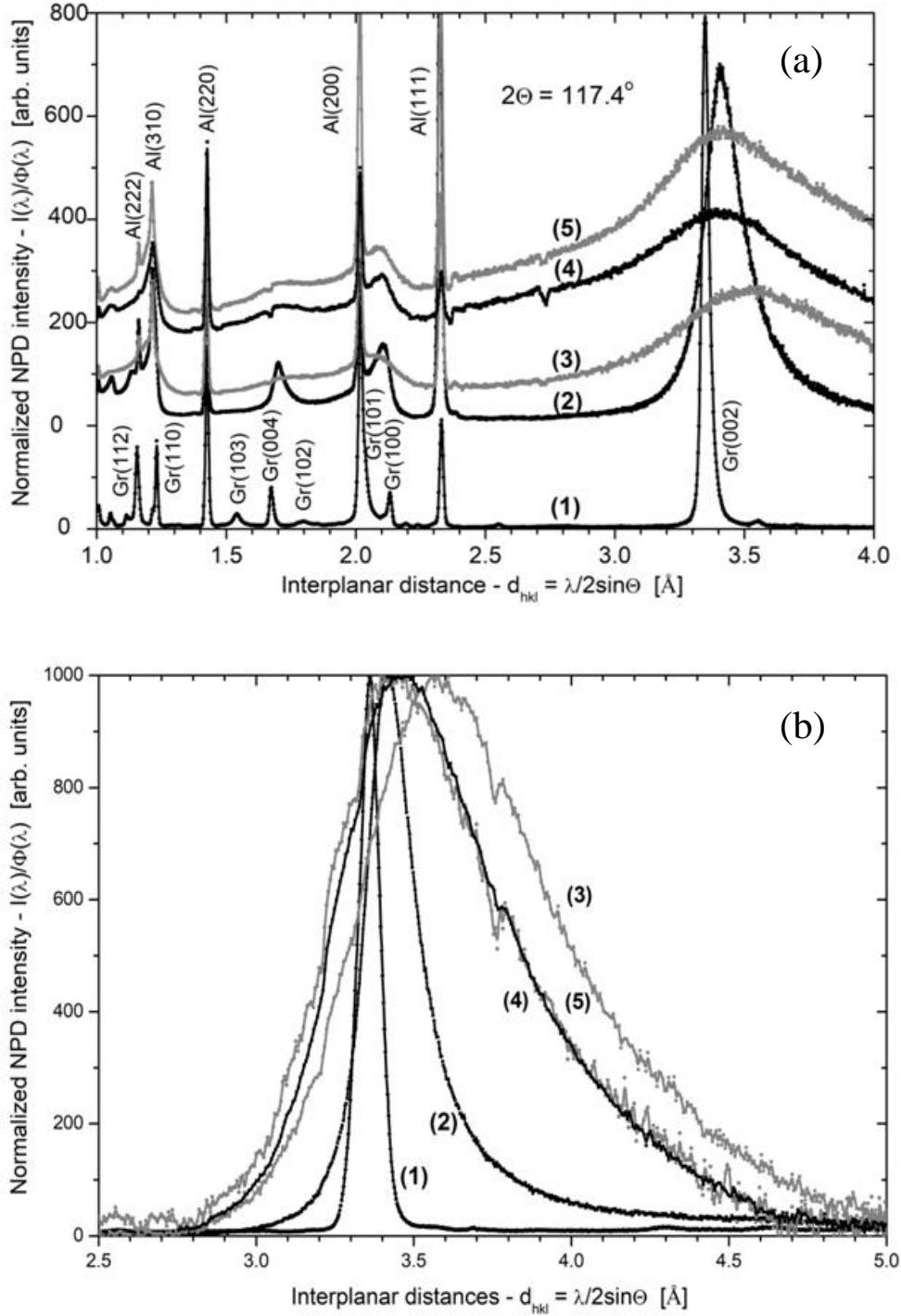

**Figure 1.** NPD panorama view (a) and Gr(002) reflexes (b) of *as prepared sp$^2$*- amorphous carbons and graphite: 1 – Gr (20K); 2 – CB624 (6K); 3 – CB632 (6K); 4 – ShC (20K,) 5 - AntX (6K), scattering angle $2\Theta=117.4^0$.

CSR length $L_{CSR}$ are inversely connected: $B = k\lambda / L_{CSR} \sin \Theta$, where $k$ is, strictly speaking, an undetermined parameter while $\lambda$ and $\Theta$ are the neutron wave length and scattering angle. When the diffraction study of a set of samples is performed under the same conditions, it is possible to take one of the samples as a reference one and to determine $L_{CSR}$ of the remaining samples addressing to that
5

of the reference. In our study, $L_{CSR}^{ref}$ is attributed to crystalline graphite and constitutes ~20 *nm* along both *c* and *a* directions [35]. Therefore, $L_{CSR}$ of the studied amorphics can be determined as

$$L_{CSR} = (B_{ref}/B)(\lambda/\lambda_{ref})L_{CSR}^{ref} \qquad (4)$$

Applying the relation to the data presented in Fig. 1b, we obtain that $L_{CSR}^c$ for ShC and AntX is practically the same and corresponds to 2.5 nm thus evaluating the thickness of BSU stacks covering ~7 graphene layers while BSU stacks of CB624 and CB632 are 7.8 nm and 2.2 nm, respectively, corresponding to 23 and 6 graphene layers. In the latter case, upshifting the peak maximum is significant as well. Apparently, both the width of the peak and its maximum position may be influenced by pronounced turbostratic arrangement of BSU in stacks. The problem needs a further examination.

As pointed in [36], $L_{CSR}^a$, which determines the lateral dimension of graphene-like BSUs, is characterized by the width of Gr(110) reflex located at 1.24Å in graphite. As seen in Fig. 1a, the corresponding reflexes of the studied amorphics, besides CB624, are markedly downshifted and broadened pointing to size contraction in this direction. However, since intense Al(310) reflex of cryostat does not allow a reliable extraction of the peak from NPD data, the only attempt has been undertaken for ShC [19]. According to the evaluation, $L_{CSR}^a \sim 1.9 L_{CSR}^c$. Taking into account a possible turbostratic BSU arrangement within stacks, the obtained $L^a$ ~4 nm can be considered as the upper limit of the lateral size of graphene-like sheets forming BSUs of the studied amorphics while the size of individual BSU might be significantly less. Evident similarity of the NPD pattern in this region for ShC, AntX, and CB632 allows concluding that the lateral dimensions of the three amorphics are of the same order. The lateral BSU size in CB624 is obviously much bigger.

A comparative analysis of the NPD plottings has allowed revealing the following structural regularities related to the studied ACs.
1. The selected natural and synthetic amorphics do belong to graphitic carbons, the main structural fragments of which consist of stacks of graphene-like BSUs albeit of different structure.
2. The coincidence of Gr(002) reflex maxima positions for natural amorphics at 3.47 Å evidences a similarity of the structural composition of their BSU stacks consisting of 7 graphene layers.
3. The difference in the maxima positions of Gr(002) reflex for CB632 and CB624 at 3.57 Å and 3.40 Å, respectively, as well as drastic difference in the peaks width point to remarkably different BSU stacks consisting of 6 and 23 graphene layers. The lateral dimensions of the relevant BSUs differ drastically as well constituting structure concerning not only their linear dimensions, but turbostratic disordering as well that is evidently large in the case of the CB632 sample.

**Inelastic Neutron Scattering**

There were two experimental sessions in due course of the project fulfillment differing by cryostats used and temperature regimes. Figure 2 presents TOF INS spectra obtained in the course of the two sessions, recorded at T=20K and 6K, respectively. The spectra are summed over 15 scattering angles and normalized per 10 hours of exposition time. Plottings presented in Fig. 2a are related to ShC at different attempts and lay the foundation for a forthcoming normalization of the spectra intensity providing a quantitative comparison of the obtained results. Spectra in Fig. 2b are related to *wet* and *dry* AntX and CB632. The spectra of *wet* CB624 and *as prepared* graphite of the same mass are at the level of the instrumental background and cannot be distinguished from each other. In contrast, all other amorphous samples scatter neutrons quite intensely, indicating to be evidently hydrogen-enriched. To reveal the kind of this enrichment and making it possible to be quantitatively analyzed, the recorded spectra were exempted from the background. Thus treated TOF spectra, normalized per 10 gram of mass, are shown in Fig. 3. The spectra form the ground for a quantitative comparison at an acceptable level of accuracy. As seen in the figure, each amorphics can be characterized by a three-component set of spectra related to *wet* and *dry* samples as well as to adsorbed water that is presented by the difference of the two first spectra. For mesoporous amorphics similar to the studied samples, the presence of adsorbed water is not a surprise. Much bigger surprise is the fractional contribution of the water spectrum in the spectra of *wet* samples. As seen in Fig. 3, this fraction constitutes ~0.3 and ~0.5 in the case of AntX and ShC, respectively, leaving a considerable contribution for scattering from both *dry* samples and evidencing that the carbon cores of the species are significantly hydrogen enriched as well. In contrast, the water contribution dominates in the spectrum of *wet* CB632 leaving only ~10% of the total intensity to the dry sample, the TOF INS spectrum of which is practically identical to that of graphite (see Fig. 2). This finding convincingly evidences that the carbon core of the sample is not markedly hydrogenated. This conclusion in the case of CB624 is even more categorical since, as was said earlier, the TOF INS TOF spectrum of the latter *wet* sample is practically identical to spectrum (2) in Fig. 2a.

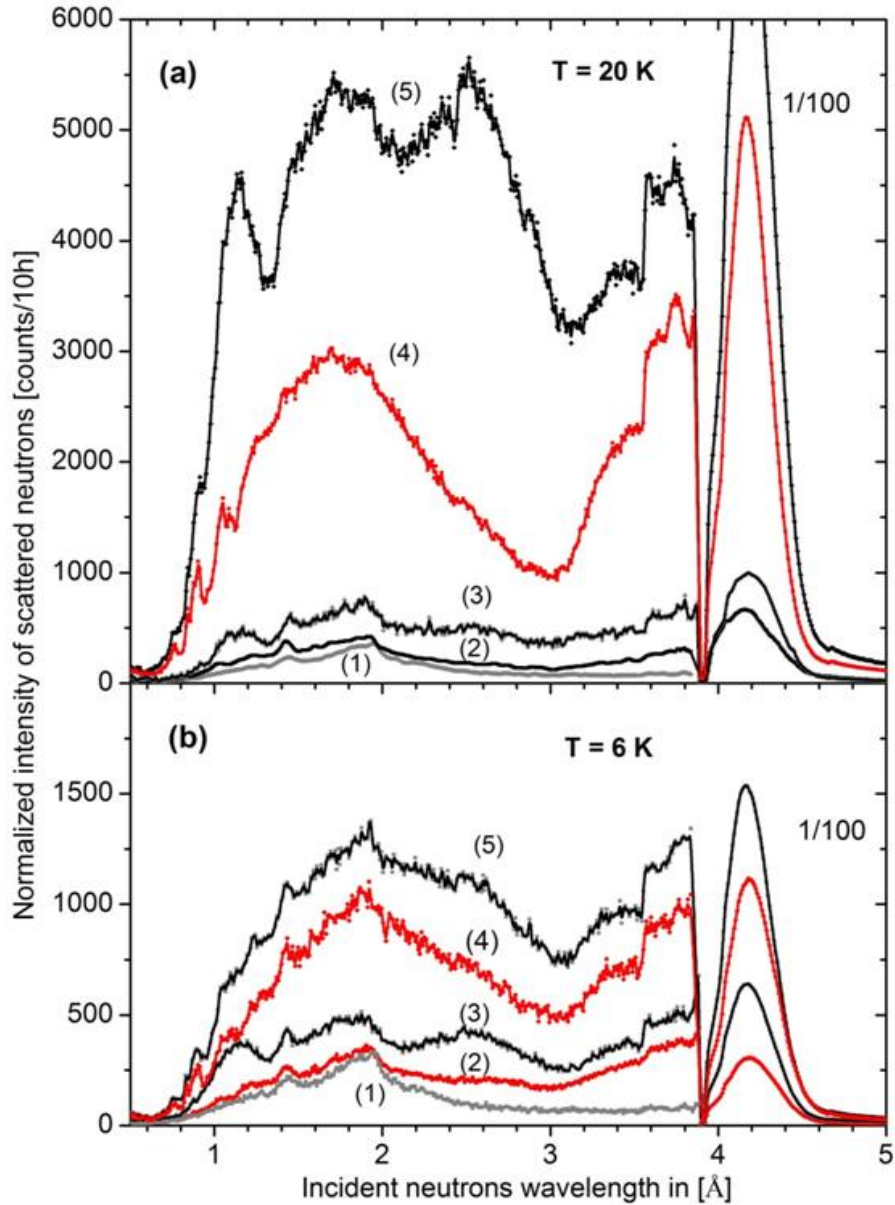

**Figure 2.** TOF INS spectra of as prepared (*wet*) and dried (*dry*) amorphous carbons:
(a) (T = 20K). 1. Al cryostat-1; 2. Gr; 3. ShC (*wet*,10g); 4. ShC-(*dry*, 96g); 5. ShC (*wet*,100g).
(b) (T = 6K). 1. Al cryostat-2; 2. CB632 (*dry*, 9.7g); 3. CB632 (*wet*,10g) 4. AntX (*dry*,10g); 5. AntX (*wet*,10g). Intensity of elastic peaks is hundredfold reduced.

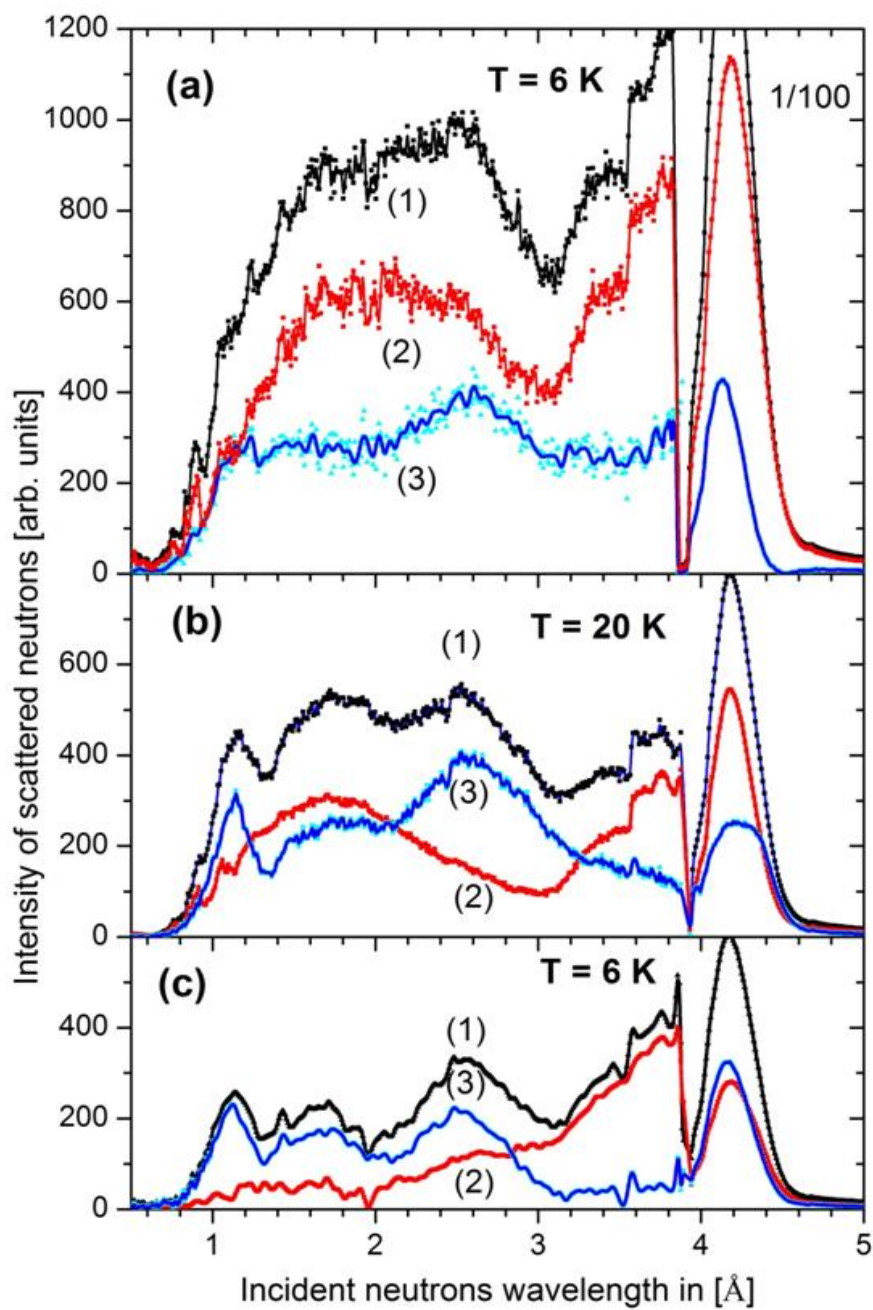

**Figure 3**. Normalized TOF INS spectra of ACs after subtraction of linear background: (a) - AntX; (b) – ShC; (c) - CB632. Digits mark spectra of *wet* (1) and *dry* (2) samples as well as of adsorbed water (3).

A comparative analysis of the obtained TOF INS spectra has allowed revealing the following regularities related to the studied amorphous carbons.

1. The INS intensity from natural amorphics presented by *wet* AntX and ShC is quite high and comparable while that one from synthetic amorphics varies from barely measurable (CB632) to very low (CB624).
2. The INS spectra of all the amorphics (besides CB624) are composed of two contributions related to adsorbed water and the solid carbon cores.
3. Both INS contributions of the natural amorphics are comparable evidencing hydrogen enriched nature of their carbon cores.
4. Water contribution to INS spectra of synthetic amorphics either dominates (CB632 - good adsorbent) or is not fixed at all (CB624 - poor adsorbent) leaving the contribution of their carbon cores at very low level in all the cases and evidencing non-hydrogeneous nature of the latter.

To make clear the chemical content of the studied amorphics carbon cores, we turn to energetic plottings of the recorded spectra in terms of GVDOS - $F(\omega)$ in the form of Ex. (2). When performing the treating, we ignored multiple scattering and suggested one-phonon character of the scattering in form of Ex. (1). The former is supported by specially optimized shape of the sample container while the latter was provided by low temperatures.

## IV.  RESULTS AND DISCUSSION

**GVDOS experimental spectra**

Experimental GVDOS spectra, corresponding to TOF INS ones presented in Fig. 3, are shown in Fig. 4. As seen in the figure, the spectra preserve three-component constitution, the main features of which are described in the previous section. Since experimental data were normalized per mass, per time, per flux, the spectra features can be compared at a quantitative level and directly attributed to those of the relevant BSUs. Further consideration will be divided in two parts devoted to spectra of adsorbed water and the ACs cores. We first proceed with the first topic and then turn to the structure-atomic compositions of BSU responsible for the core spectra. The BSUs structural properties concerning size and shape should follow to the performed neutron diffraction study, while those concerning chemical content should fit CVDOS spectra (2) in Fig. 4.

**INS from adsorbed water**

Water is willingly adsorbed in both natural and synthetic amorphous carbons, which is clearly seen in Fig. 4 and is evidently caused by their porous structure. The state of art of the retained water is determined by two facts, namely, the pore size and chemical content of the pores inner shells. Previous INS study allowed considering the water behavior in ShC pores [18, 19]. Since water spectra of the studied samples are much in common, we shall use the ShC results as a reference when analyzing water behavior in AntX and CBs. As shown for ShC, water molecules interacts with pores shells quite weakly, which allows behaving them as free molecules. This thesis is directly confirmed in the current study on the example of CB632. Aldrich's carbon blacks are known as perfect water adsorbent, which is provided by a considerable porosity of the samples, on the one hand, and hydrophilicity of the cores inner shells, caused by their oxygen-enriched structure, on the other. The first factor allows performing experiment with calibrated dozing of adsorbed water. Figure 5 presents a set of TOF INS spectra obtained under the study. Experimental spectra 4-6, attributed to adsorbed water in CB632, were obtained by extracting spectrum 2, related to *dry* CB632 sample, from spectra related to this sample with additionally added water. Spectrum 7 is related to CB624 and was obtained after exclusion of background spectrum 1 from this sample spectrum with added water. The spectrum occurred to be an independent support of the validity of quantitative correlation between the spectra intensity and adsorbed water mass. Thus, a good coincidence of spectra 6 and 7 may evidence that the scattering from 2.5 ml of water in CB624 is equivalent to the scattering form 2.0 ml added water in CB632 plus adsorbed water in the sample. Due to the observed equivalence, the quantity of adsorbed water should be ~0.5 ml. As follows from independent analytical study [16], the quantity constitutes 1.5 ml, which is a rather good fitting for so different tools of testing.

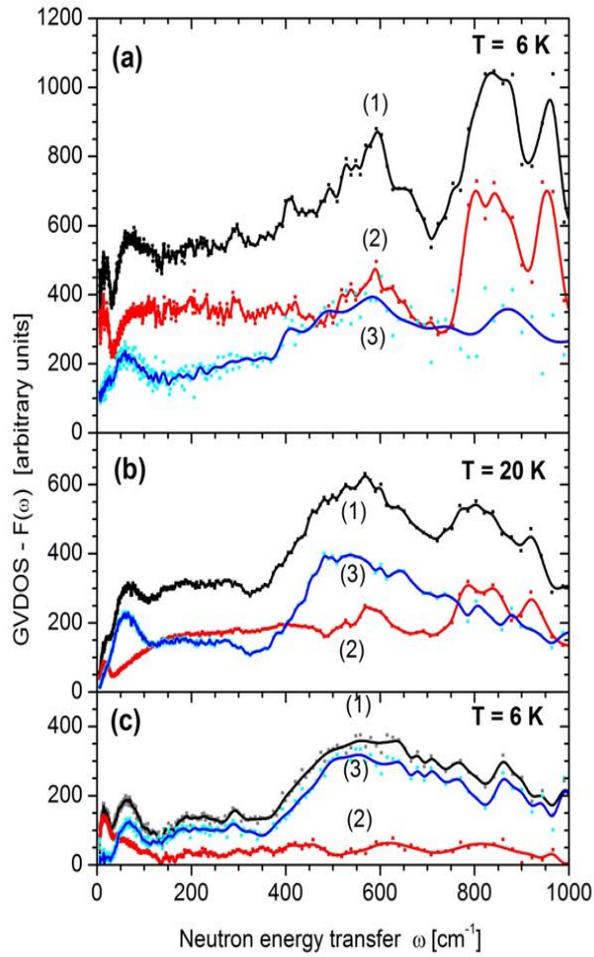

**Figure 4.** The GVDOS-$F(\omega)$ spectra of AntX (a); ShC (b); CB632 (c). Digits marking: (1) *wet* and (2) *dry* samples; (3) adsorbed water.

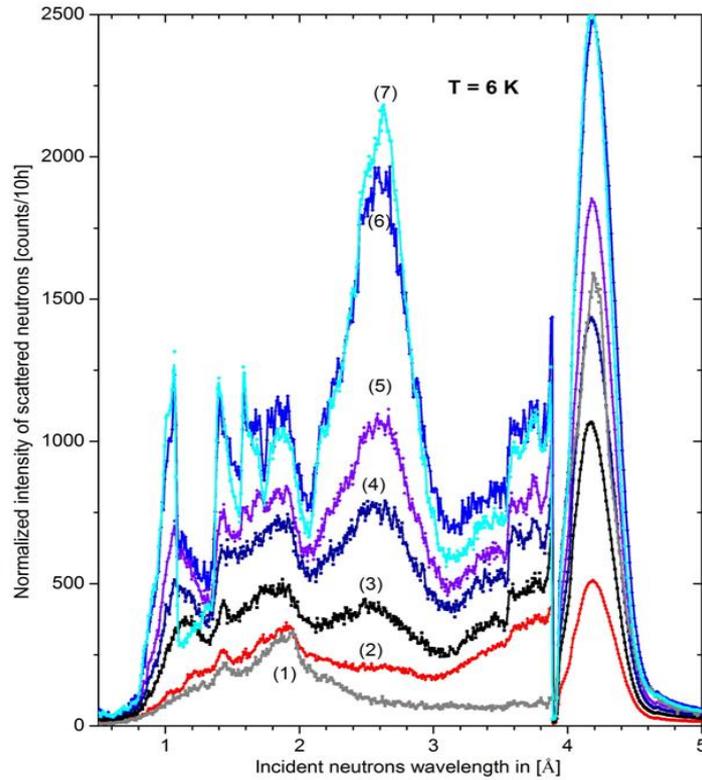

**Figure 5**. TOF INS spectra of adsorbed water in 100 g of carbon black CB632: (1) *Al-cryostat*; (2) *dry* and (3) *wet* pristine samples. (4) – (7) spectra of water additionally added to pristine samples: namely, 0.5 ml (4), 1.0 ml (5), and 2.0 ml (6) added to sample *wet*; 2.5 ml added to 100 g of CB624 *wet* (7).

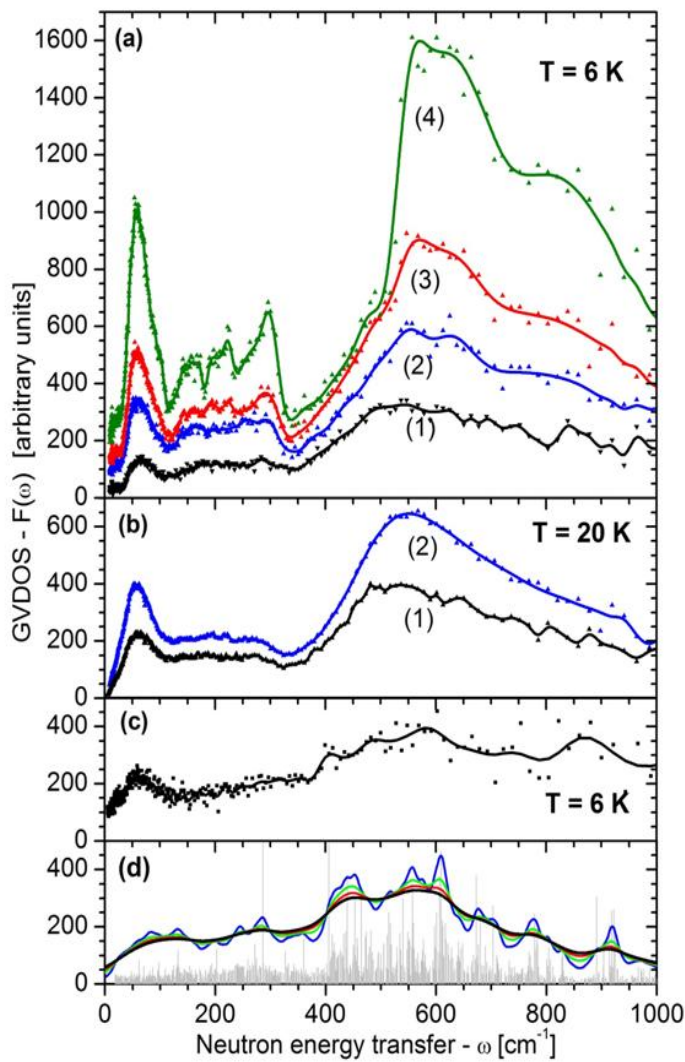

**Figure 6.** Experimental GVDOS –$F(\omega)$ spectra of adsorbed water in amorphous carbons. a. CB632: (1) – (4) 0; 0.5; 1.0 and 2.0 ml of added water, respectively. b. ShC as prepared *wet* (1) and oxidized ShC(ox) (2). c. AntX. d. Calculated spectra of four-hydrogen-bond configured molecules of retained water, convoluted with Lorenzian of different half-width from 20cm$^{-1}$(bleu curve) to 80 cm$^{-1}$ (black curve). See details in Ref. [39].

Thus obtained TOF INS spectra were converted into GVDOS ones, presented in Fig. 6a. As seen in the figure, the spectra are continuously transformed towards the spectrum typical to ice when the water amount increases. As well known (see [18, 19] and references therein), the ice INS spectrum at NERA spectrometer is composed of the contribution of three main groups of vibrations, namely, (I) hydrogen bonds (HBs) bendings at 59 and 150 cm-1, (II) HB stretchings at ~224 and ~296 cm$^{-1}$ as well as (III) librational, or hindered translational and rotational, modes forming a broad band in the region of 600-1200 cm$^{-1}$. Band I is additionally convoluted with the low-frequency Debye phonon-like acoustical contribution. The latter modes are present in water because of intermolecular hydrogen bonds (HBs) that are formed by each water molecule surrounded by a quartet of other ones. Consequently, the molecule motion is quite collective, due to which it greatly depends on external constrains.

In experiment, results of which are presented in Fig. 6a, constrains are implemented by the limited size of pores, on the one hand, and different volume density of water, controlled by its concentration in the pores, on the other. Actually, the only ice mode positioned at ~56 cm$^{-1}$ (band I) is retained in the $F(\omega)$ spectrum up to the lowest concentration while all the bands in region II reveal clearly seen flattening when concentration goes lower. Analogous spectral modification takes place with respect to the ice librational modes forming broad band III. The band is provided with water molecule rotations around three symmetry axes whose partial contribution determines the band shape. As shown by detailed studies [37], the modes conserve their dominant role in the INS spectra of retained water, albeit are downshifted, when water molecules are coupled with the pore inner surface via hydrogen bonds. This very behavior is characteristic for the $F(\omega)$ spectra in Fig. 6a. According to [37], the downshift value of the red edge of band III from 550 cm$^{-1}$ in free ice may be used to evaluate the pore size. Thus, its reducing to 230 cm$^{-1}$ in the case of CB632 as well as of ShC (spectrum (1) in Fig. 6b) and AntX (Fig. 6c) points pore size of a few nm. This estimation perfectly correlates with the linear size of 1.5-2.5 nm of BSUs themselves as well as of their stacks in the studies ACs since the pore size correlates with the linear size of elements by which they are constructed. It should be mentioned as well the evidently similar pattern of the water spectra, related to comparative adsorbed water amount in all three cases (see spectra 1 in Figs.6a-6c).

Experimentally determined, the amount of adsorbed water in the studied as prepared *wet* ACs forms a series of 4 wt%, 2 wt%, 1.5 wt%, and 0.6 wt% related to ShC, AntX, CB632, and CB624, respectively [16]. Since INS study does not reveal the presence of adsorbed water in CB624, the amount of 0.6wt% can be considered as either equal to or less than the limit of the water amount sensitivity of the used technique. Anyway, with this figure in mind the estimation of about 0.5 wt% of adsorbed water in CB932 on the basis of the coincidence of spectra 6 and 7 in Fig.5 seems to be more reasonable. As to the series in the whole, as seen in Fig.6, it is convincingly justified experimentally: spectrum 1 of ShC is the most intense, then follows spectrum 1 of AntX approximately twice lower by intensity and finally is spectrum 1 of CB632 by one fourth time less than AntX spectrum. Evidently, the closer is the water amount to the sensitivity limit, the more is deviation of the spectrum intensity from the 'linear' dependence on the water amount.

It should be noted that the total amount of confined water is evidently governed by two main factors, namely: the core size and the content of oxygen atoms in the inner shell of the pores. As was determined empirically [16], all the studied ACs represent a complex agglomeration of framed oxyhydride graphene molecules (BSUs), different by linear size and averagely stable by the oxygen content for each AC while varied by the hydrogen content and hydrogen-oxygen atomic bonding concerning the molecules circumferences. The first feature mandatory affects the pore size due to which the latter forms the following series CB624>>>ShC>AntX>CB632, which is based on the structure data of the studied ACs [16]. According to the series, the pore size of CB624 differs tenfold from that one of other ACs. So big pore size is, certainly, quite unfavorable factor for the water retention in the species since the latter is usually characteristic for rather small pores. This feature explains the poor quality of CB624 as water adsorbent while that one is significantly improved in the small-pore species and provides the subordination of the confined water amount to the series pore-size regularity. The second feature determines the pore hydrophilicity and strictly depends on the BSUs chemical composition. A detailed investigation [16] has revealed that the studied amorphics belong to different subclasses of $sp^2$ amorphous carbon differing by the composition of oxygen-containing groups representing the main feature of the relevant BSUs. Thus, shungite carbon was attributed to the "C=O" subclass, the main feature of which is provided by carbonyls, acid anhydrides and *o*-quinones. Antraxolite was attributed to the "O=C-O-C" subclass, for which lactones, pairs of lactones, aggregated cyclic ether with lactone as well as partially hydroxyls and hydroxy pyrans play the role. Cyclic ester, aggregated cyclic ester and aggregated cyclic ether with lactone are main participants in the case of "C-O-C" carbon blacks. A set of selected representatives of these subclasses on the top of Fig. 7 are characterized by the least total energy and, thus, the largest coupling energy between the molecules carbon core and the relevant addends [16]. A naked (5,5)NGr rectangular graphene sheet with five benzenoid units along both armchair and zigzag edges lays the foundation of these models. Its lateral dimensions cover the space of 1.5x1.5 nm$^2$ in van-der-Waals diameters. The models were suggested based on chemical content of the species. Evidently, the presence of four oxygen atoms in the BSU framing area becomes a governing factor concerning the retention of water molecules inside the pores of each studied AC. Therefore, any marked difference in the pore structure concerning both the core size and shape might be revealed by the INS spectrum of adsorbed water. Thus, spectrum 2 in Fig. 6b is related to ShC(ox) subjected to natural additional oxidation in the course of strong hydrothermal influence (Maksovo deposit, Karelia). In comparison with ShC, it is characterized by higher total pore volume, higher open porosity and lower pH of suspension (4.0-4.5) [38] which points increasing of the number of oxygen containing groups. Evidently, changes in the sample porosity and chemical composition of the BSU framing are well consistent with the growth of the adsorbed water amount up to 5 wt%, which is well seen in Fig. 6b. Additionally, INS study revealed that the changes do not touch the core structure of the ShC(ox) BSUs, the core spectrum of which is identical to that one of ShC.

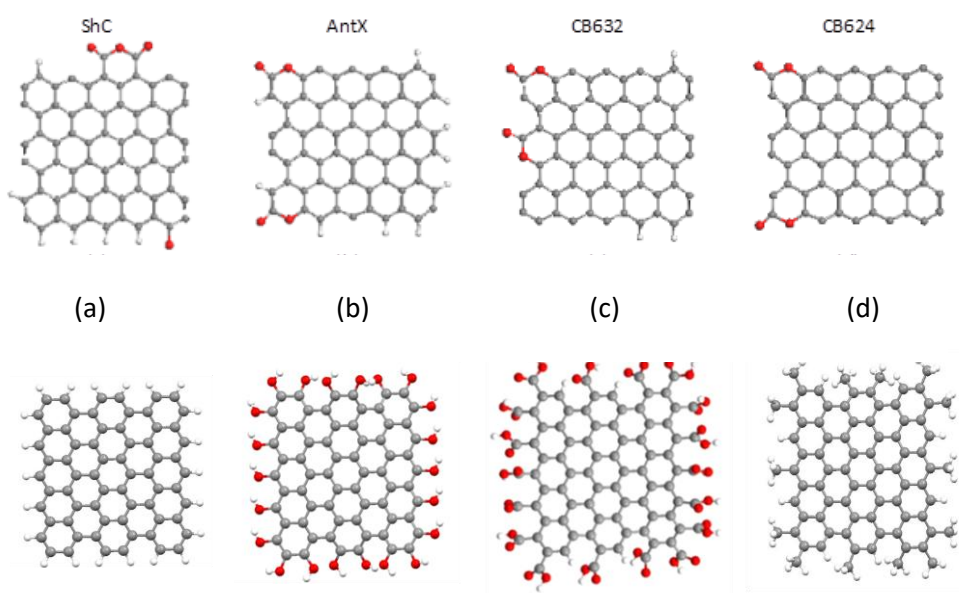

**Figure 7.** Equilibrium structures of molecular BSU models. *Top*: C$_{68}$O$_4$H$_6$ model of "C=O" shungite carbon (a), C$_{64}$O$_4$H$_{10}$ model "O=C-O-C' antraxolite (b), C$_{64}$O$_4$H$_3$ (c) and C$_{64}$O$_4$ (d) models of "C-O-C" carbon blacks CB632 and CB624, respectively. UHF calculations. Adapted from [16]. *Bottom*: Standard "H-models": C$_{66}$H$_{22}$ (a); C$_{66}$(OH)$_{22}$ (b); C$_{66}$(COOH)$_{22}$ (c); C$_{66}$(CH$_3$)$_{22}$ (d). DFT calculations. Red, gray and white balls mark oxygen, carbon and hydrogen atoms, respectively.

Computationally, the adsorbed water was previously analyzed once confined between layers of graphene oxide [25, 39]. Presented in Fig. 6d is a set of calculated $F(\omega)$ spectra of a monolayer of water molecules coupled by fours hydrogen bonds with the surrounding. The spectra differ by the half-width of the convolution functions. As seen in Fig. 6, the L80 spectrum (black curve) is quite similar to experimental spectra of all *as prepared* samples thus convincingly pointing to a mono-layer character of water molecules disposition within the pores of the studied amorphics.

**BSUs models and GVDOS calculated spectra of the amorphous carbon core**

According to a comparative study of spectra (2) in Fig.4, the following features are characteristic for the cores of the studied ACs: 1. The enough-intense core spectra of AntX and ShC are well similar by shape, whilst somewhat different in details, but differing by intensity about twice in the favor of AntX; 2. The low-intense core spectra of CB632 and CB624 are similar to each other as well to the reference graphite spectrum, while nevertheless the former is slightly more intense than the other two. Attributing the discussed findings to peculiarities of the relevant BSUs, one can evidently conclude that BSUs of natural amorphics are significantly hydrogenated while those of the synthetic products are hydrogen-poor if not fully non-hydrogeneous. At a qualitative level, the findings are in general agreement with the model structures of the amorphics BSUs presented on the top of Fig. 7 due to which the latter might be quite convincingly used to present detailed features of the GVDOS – $F(\omega)$ spectra of the related amorphics.

However, the computation of $F(\omega)$ spectra of the discussed models is encountering the problem. The matter is that empirically evidenced BSUs of the studied amorphics are large molecular stable radicals, which is exhibited, in particular, by models shown in the figure [16]. Solving a dynamic problem for large molecular radicals is a difficult task outside the currently widespread calculation methods based mainly on the spin nondependent DFT approach suitable for the consideration of closed-shell electronic systems. Open-shell electronic configuration of radicals requires a different approach so that the information concerning the spin-dependent dynamic problem solving in the case of large molecular radicals is rather scarce [40, 41]. At the same time, the closed-shell approximation, realized as an all-electron numerical method for solving the local density functional for polyatomic molecules [29] and implemented in DFT-based DMOL3 program [30], is widely used for the calculation of vibrational spectra of graphene-based molecules (see [25, 39] and references therein) suggesting rather good results for comparison with empirical data. However, this approach works not in all cases. Concerning the above set of models in Fig. 7, program DMOL3 does not allow reaching the equilibrium position of these radical molecules, thus making impossible solving the dynamical problem. Due to this circumstance, in the current study the open-shell UHF approach, implemented in the AM1 semiempirical tool CLUSTER-Z1 [31], was suggested to solve spin-dependent dynamic problem of highly radicalized graphene molecules. Simultaneously a few less radicalized molecules, representing graphene flakes with totally terminated edge atoms, were successfully considered by DMOL3 program. These molecules, shown on the bottom of Fig. 7, form a set of particular models produced "H-standard" INS spectra of (5,5)NGr-based BSUs. As seen in the figure, (5,5)NGr molecule represents the carbon core of all the molecules, edge atoms of which are fully terminated by H, OH, COOH, and $CH_3$ units.

**Comparison of experimental and calculated spectra**

Virtual "H-standard" $F(\omega)$ spectra are presented in Fig. 8 in the frequency scale accessible for the INS measurements at NERA spectrometer. Panel (a) in the figure is related to the model molecules as a whole while panel (b) exhibit the contribution provided by the molecules carbon and oxygen atoms only. The panel comparison clearly reveals that the INS intensity from the heavy atoms is negligibly small. This is due to that the neutron scattering cross section of the hydrogen atom is more than one order of magnitude greater than those of the other atoms (C, O, N, S . . . ). Therefore, the carbon matrix is largely transparent for neutrons so that its empirical INS spectra are provided by the vibration involving the hydrogen atoms. The discussed "H-standard" spectra are highly characteristic, which allows attributing particular spectra features to specific atomic groups containing hydrogens. Thus, the peak at ~850 $cm^{-1}$ should be attributed to the out-of-plane deformational vibrations provided by C-H groups: it presents a dominant feature in the spectrum of $C_{66}H_{22}$ molecule and is clearly seen in the spectrum of $C_{66}(CH_3)_{22}$ molecule as well. Alongside with a dominant feature at ~150 $cm^{-1}$, this peak is a characteristic mark of the presence $CH_3$ group in the latter case. Actually, the peaks were observed empirically when the chemical modification of the graphene molecule circumference occurs in the presence of propyl alcohol [25]. Continuing the description of H-spectra it becomes evident that strong peak at ~650 cm$^{-1}$ followed with less intense peak at ~80 cm$^{-1}$ in spectrum of $C_{66}(COOH)_{22}$ molecule are characteristic marks of COOH units while those ones at ~350 cm$^{-1}$ and ~550 cm$^{-1}$ in the spectrum of $C_{66}(OH)_{22}$ evidence the presence of hydroxyls. Comparison of the discussed H-standard spectra with experimental ones presented in Fig.4 allows reliably concluding that hydrogen atoms are incorporated in the cores of the studies ACs by the formation of C-H bond with the BSU edge atoms thus confirming previous conclusions made for ShC [18-21].

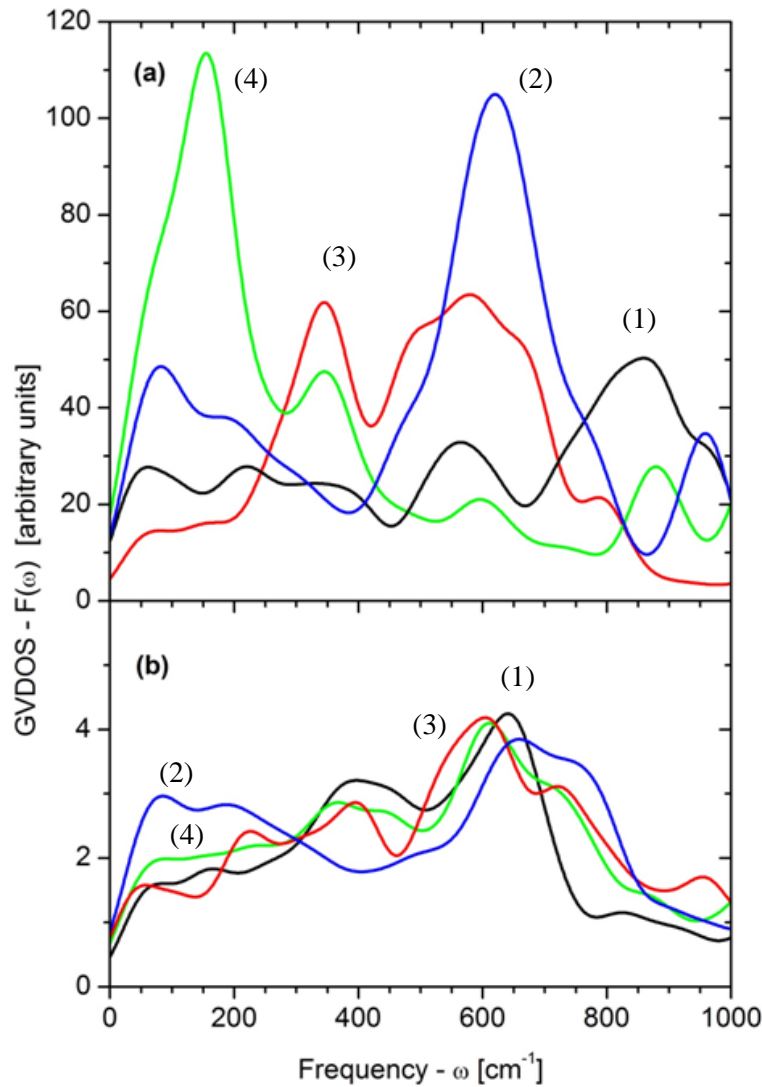

**Figure 8**. DFT simulated "H-standard" GVDOS – $F(\omega)$ spectra: models $C_{66}H_{22}$ (1); $C_{66}(OH)_{22}$ (2); $C_{66}(COOH)_{22}$ (3); $C_{66}(CH_3)_{22}$ (4) (see Fig. 7(bottom)). Scattering from C, O, H atoms (a) and C, O atoms only (b). The spectra are convoluted by Gaussian of 80 cm$^{-1}$ half-width.

It should be mentioned that the INS spectra image is not spectrometry standard as, say, IR or Raman spectra. The INS fixed image strongly depends on both the neutron source and the way of the neutron flux registration implemented in the spectrometer in use. Thus, the spectra in Fig.8 are typical for NERA spectrometer at IBR-2 high flux reactor. Nevertheless, the characteristic H-standard peaks related to CH, CH$_3$, OH, and COOH units at ~850 cm$^{-1}$, ~150&850 cm$^{-1}$, ~350&550 cm$^{-1}$, and ~80&650 cm$^{-1}$, respectively, remain equally characteristic for other spectrometer such as IN1, TOSCA, TFXA, MARI and so forth as well (see INS data presented in monograph [42] and references therein).

Model BSU structures on the top of Fig.7 exhibit the presence of such bonds in the studied ACs based on the analytical investigation [16]. The corresponding virtual GVDOS – $F(\omega)$ spectra of these models are shown in Fig.9. The spectra were normalized per summary intensity of the pristine δ-functions that constitutes 1279, 755, 440, and 23,5 arb. units from top to bottom, respectfully, thus providing quite visible spectra for AntX, ShC, and CB632 and pointing to practically nil spectrum of CB624. Inset in panel (c) of the figure demonstrates practically linear dependence of the spectrum total intensity on the number of hydrogen atoms. The 'nil-level' scattering from heavy atoms provides INS spectrum in panel (d), whose intensity is manytensfold less than of those presented above. Generally, the serial regularity perfectly suits the experimental ones as follow from the comparison of spectra 2 in Fig. 4. The only inconsistency concerns the spectrum of CB632 that empirically is much less intense that in computations. The matter is that the amount of hydrogen in the sample determined analytically constitutes 0.3 wt%, which is twice less than that of ShC [16]. Accordingly, it was accepted the number of hydrogen atoms in the relevant model to be twice less than for ShC and constitute 3. However, as seen in Fig. 4, the quantity is evidently over evaluated since the relative intensity of the ShC spectrum is not 2 but about 4-5 times higher than that of the CB632 spectrum. It means that the hydrogen amount in the latter is less that 0.3 wt%. At the same time, this value is at the level of statistical nil of elemental analysis and cannot be considered as too accurate. As turned out, INS study is more sensitive in the case and following the inset in panel (c) and the earlier discussed intensity ratio one should conclude that N≤1, which might correspond to 0.1 wt% of the total amount of hydrogen in CB632 sample or even less.

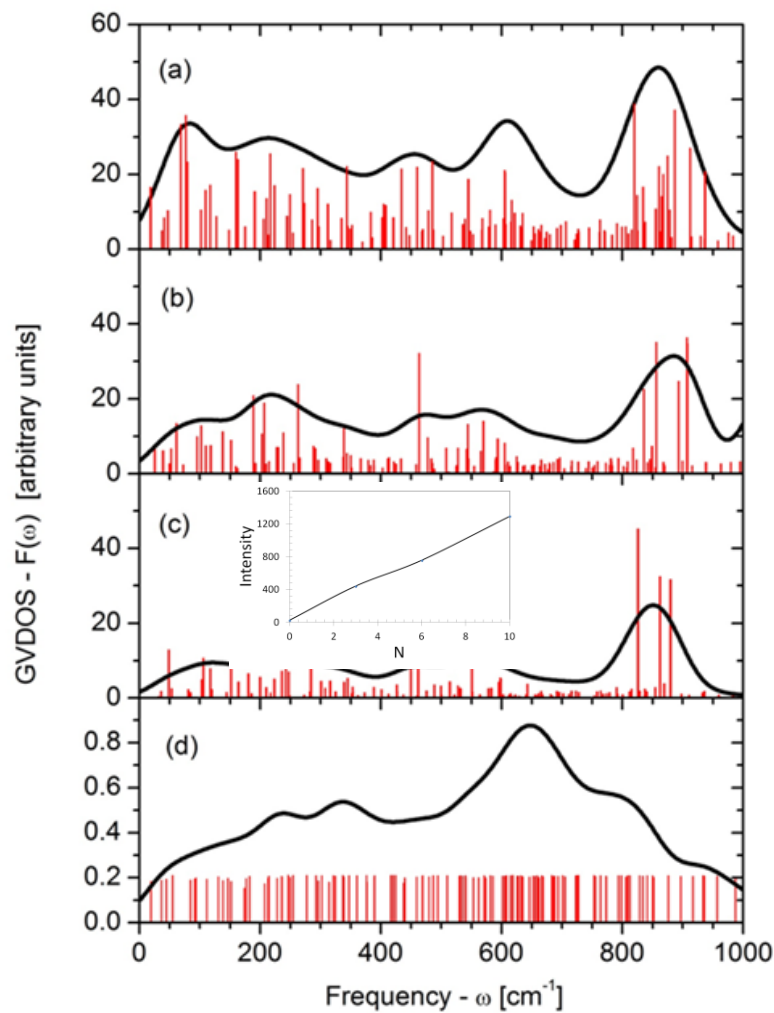

**Figure 9**. UHF simulated GVDOS – $F(\omega)$ spectra for BSU models $C_{68}O_4H_6$ (a), $C_{64}O_4H_{10}$ (b), $C_{64}O_4H_3$ (c), and $C_{64}O_4$ (d) (see Fig. 7 (top)). Bars represent sets of δ-functions from Ex.(2). Black curves are results of these functions convolution by Gaussian of 80 cm$^{-1}$ half-width. *Inset*: total intensity of the spectrum vs. the number of hydrogen atoms.

Coming back to Fig.9, it should be noted that the spectra of the first three ACs are well similar by shape as well as to spectrum of $C_{66}H_{22}$ molecule in Fig.8 thus evidencing C-H character of the hydrogen atom attaching to the molecules edge atoms and direct dependence of the spectra intensity of the number of the formed bonds. The main H-standard feature at ~850 cm$^{-1}$ is distinctly observed in the ShC and AntX spectra while rather smoothed in the CB632 case. The main contribution below 800 cm$^{-1}$ is provided with vibrations of heavy atoms, but drastically enhanced due to 'riding effect' caused by the participation of hydrogen atoms in the eigen vectors of the relevant vibrational modes [43]. This effect allows revealing changes concerning these data vibrations caused by structural reconstruction of the models. The changes are clearly seen in δ-function distributions presented in Fig.9 while unavoidable broadening of real spectra greatly smoothes the effect. And yet, the difference in the structure of spectra (2) in Fig.4 in the region below 800 cm-1 may be confidently accepted as evidencing the difference in the heavy atom structure of the studied ACs.

## V. CONCLUSION

The current study has shown one more time that both neutron diffraction and inelastic neutron scattering are powerful techniques to investigate intimate details of nanostructured materials concerning the size, structure and chemical composition of their basic structure units. Particular sensitivity to hydrogen (protium) contribution makes the INS techniques to be able to reveal the nucleus (atom) contribution at the level of ~0.1wt%/ which cannot be surpassed by any other modern analytical techniques. This exclusive feature makes hydrogen marking of the materials a strongly distinguishing characteristic. Stemmed from this, the available data on the INS study of different carbons, including results obtained in the current study, allow suggesting a clear their attribution to different subclasses. Thus, natural $sp^2$ amorphous carbons, namely, shungite carbon [18, 19, 21, 25, 44, 45], antraxolite, and coals [46] can be evidently characterized as hydrogen-enriched while described by a distinct C-H mode of H-standard INS spectra similar to those presented in Figs. 4a and 4b. The second subclass covers such industry engineered products as different carbon blacks, including furnace black, gas black, activated carbon [47, 48], diesel soot [49] and others (a profound information can be found in monograph [42]). All the products are poor-hydrogen-enriched and their INS spectra present the spectra of carbon cores slightly enhanced by riding effect as is typical to spectrum 2 in Fig. 4c. The third subclass concerns technical graphenes [10] which are the products of the modern high-tech industry in due course of nanoscale lamination of bulk graphite by using either mechanical milling [50] or chemical oxidation/reduction reactions [25, 51] and thermal shock [44, 52]. The hydrogen enrichment is not standardized in this case due to which the INS spectra may change from the spectrum of nanographite [50] to a pure (C-H [50]) or mixed ($CH_3$, COOH and C-H [25] as well as C-H and COOH [44]) modes of the H-standard spectra [53].

The suggested INS classification of $sp^2$ amorphous carbons is deeply rooted in the structural atomic-chemical composition of the materials basic structure units. This issue alongside with the great advantage of neutrons to vibrational spectroscopy concerning the ease of the spectrum simulation from the output of modern quantum-chemical calculations makes any attempt to match a virtual BSU model to the reality highly tempting. However, as shown in the current study, addressing INS to $sp^2$ ACs changes the attitude to the solution of dynamical problem concerning the determination of vibrational frequencies and atom displacements. The amorphics BSUs of the current case are large molecular stable radicals [16] due to which the well developed and widely used approach of the dynamic problem solving, based, for example, on the DMOL3 program [42], does not work any more and spin-dependent molecular dynamics should be considered instead. The current paper presents the first successful attempt of such consideration related to large graphene molecules and based on the unrestricted Hartree-Fock approximation. The approach facilities are quite large and it can be used in all other cases when spin-dependent molecular dynamics should be considered.

**Acknowledgements**

The authors (N.P., V.P., E.Sh.) would like to thank the «RUDN University Program 5-100» for financial support.